\documentclass[twocolumn,showpacs,prl,amssymb]{revtex4}
\usepackage{amsmath,amssymb,bbm,graphicx}

\begin{document}

\title{Fluctuations of an Atomic Ledge Bordering a Crystalline Facet}
\author{Patrik L. Ferrari}\email{ferrari@ma.tum.de}
\author{Michael Pr\"ahofer}\email{praehofer@ma.tum.de}
\author{Herbert Spohn}\email{spohn@ma.tum.de}
\affiliation{Zentrum Mathematik and Physik Department, TU M\"unchen\\ D-85747 Garching, Germany}
\date{October 17, 2003}

\begin{abstract}
When a high symmetry facet joins the rounded part of a crystal, the
step line density vanishes as $\sqrt{r}$ with $r$ denoting the
distance from the facet edge. This means that the ledge bordering the
facet has a lot of space to meander as caused by thermal
activation. We investigate the statistical properties of the border
ledge fluctuations. In the scaling regime they turn out to be
non-Gaussian and related to the edge statistics of GUE multi-matrix models.
\end{abstract}

\maketitle

\begin{figure}[t!]
\begin{center}
\includegraphics[height=7.8cm]{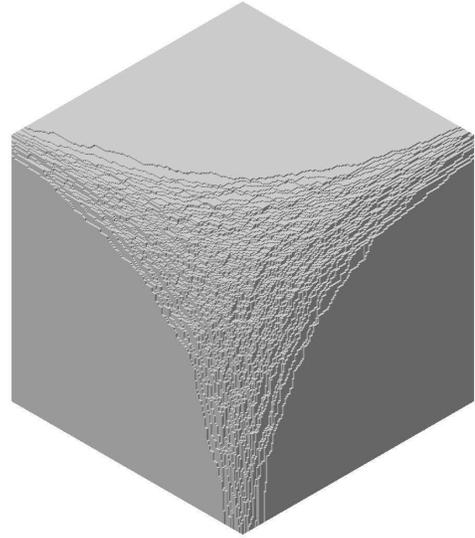}
\vspace*{-0.35cm}
\caption{\it Crystal corner viewed from the $[1 1 1]$-direction.}\label{fig1}
\end{center}
\vspace{-0.85cm}
\end{figure}

Equilibrium crystal shapes typically
consist of various flat facets connected by rounded surfaces. For a
microscopically flat facet there must be an atomic ledge bordering the
facet. This border step could be blurred because of thermal
excitations, but is clearly visible at sufficiently low temperatures
\cite{SCV,NBE,NBEB}. While in the interior of the rounded piece of the
crystal the step line density is of order one on the scale of the
lattice constant, it decays to zero as the edge of a high symmetry
facet is approached. If $r$ denotes the distance from the facet edge,
according to Pokrovsky-Talapov \cite{PT} the step line density
vanishes as $\sqrt{r}$.  Thus there is a lot of space for the border
ledge to meander, in sharp contrast to steps in the rounded part which
are so confined by their neighbors that they fluctuate only
logarithmically \cite{Sp}. The goal of our letter is to explore
quantitatively the statistics of border ledge fluctuations. To
illustrate our set-up we display in Fig. \ref{fig1} a typical
configuration from a statistical mechanics model which will be
discussed below. One clearly recognizes the
three facets as joined through a single rounded piece. Our interest is
the statistics of the uppermost ledge.

Experimentally ledge fluctuations are an elegant tool to determine
step energies \cite{SVR,NBEB}: One carefully prepares an island,
linear size $L$ and single atom height, on a high symmetry
facet. Alternatively, one sputters an undercut island. The ledge
bordering the island is well described by a random walk,
which implies fluctuations of size $\sqrt{L}$
\cite{KE}. In contrast, as can be clearly observed from
Fig. \ref{fig1}, the border ledge of a facet interacts with its neighbors
and a random walk model is not appropriate. In fact as our main
result we will establish that the border ledge has fluctuations of
size $L^{1/3}$ with a \emph{non-Gaussian} statistics.

To gain some understanding of the origin of such anomalous fluctuations let us consider the terrace-ledge-kink (TLK) model, which serves as an accurate
description of a vicinal surface, i.e.\ a crystal cut at a small angle
relative to a high symmetry crystal plane. The surface is made up of
an array of ledges which on the average run in parallel and are
separated by terraces. The ledges are not perfectly straight and
meander through kink excitations, only constrained not to touch a
neighboring ledge.  One can think of these ledges also as discrete
random walks constrained not to cross, i.e.\ with a purely entropic
repulsion. Such a line ensemble is very closely related to Dyson's Brownian
motion, in which the random walks are replaced by continuum Brownian
motions. As discussed in \cite{KM,D}, the location of the
steps at fixed random walk time $t$ have the same distribution as the
eigenvalues of a GUE $(\beta=2)$ random matrix. On this basis it is
expected that the ledge-ledge distance is governed by the GUE level
spacing \cite{EPL}. This prediction is verified experimentally
\cite{ERC}, however with deviations from $\beta=2$ which are
attributed to long range elastic forces mediated through the bulk of
the crystal and not included in the TLK model.

If in the TLK model one retains the lattice structure in the
transverse direction and makes the continuum approximation in the
direction along the ledges, then the ledges can be regarded as the
world lines of free fermions in space-time $\mathbb Z \times \mathbb
R$ \cite{BV}. The world lines are piecewise constant and have jumps
of only one lattice spacing. Consequently the transfer matrix has a
nearest neighbor hopping term and the Pauli exclusion principle
guarantees entropic repulsion in the sense that ledges never cross.

The TLK model, in the version as just explained, has no facet. The crystalline
surface has a constant average slope. Slope variations can be enforced
through a \emph{volume constraint}. For this purpose we introduce the
``occupation'' variables $\eta_j(t)$, $|j| \leq N$, $|t| \leq T$,
in the surface patch $[-N,-N+1,...,N]\times[-T,T]$: $\eta_j(t)=1$ if there is
some ledge passing through $(j,t)$, and $\eta_j(t)=0$ otherwise. In these variables,
up to an overall constant, the crystal volume is given by
\begin{equation}\label{vol}
A_{\mathrm{v}} = \int_{-T}^T\textrm{d}t \sum_{j=-N}^N j\, \eta_j(t)
\end{equation}
and volume constraint means to have an ensemble of ledges where the action
$A_{\mathrm{v}}$ is kept fixed.

\begin{figure}[t!]
\begin{center}
\includegraphics[width=8cm,height=3.5cm]{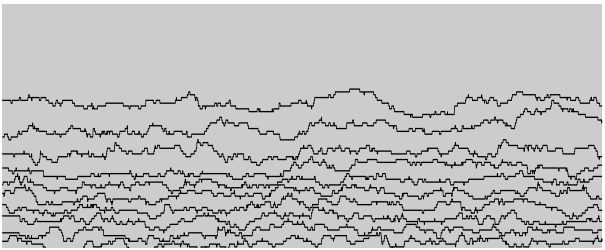}
\vspace*{-0.35cm}
\caption{\it Top lines for a TLK model with volume constraint.}\label{fig2}
\end{center}
\vspace{-.85cm}
\end{figure}

Without volume constraint the transfer matrix is generated by a
free fermion Hamiltonian
with nearest neighbor hopping \cite{BV}. Imposing the volume constraint grand-canonically
adds to the fermionic action the term $\lambda^{-1}A_{\mathrm{v}}$ with
a suitable Lagrange multiplier $\lambda^{-1}$. Thereby the nearest
neighbor hopping Hamiltonian is modified to
\begin{equation}\label{a}
H_{\mathrm{F}}=\sum_j\big(-a^\dagger_j a_{j+1}- a^\dagger_{j+1}
a_j+2a^\dagger_j a_j+\frac{j}{\lambda}a^\dagger_ja_j\big).
\end{equation}
$a_j$, resp.\ $a_j^\dagger$, is the annihilation, resp.\ creation, operator at lattice site $j\in \mathbbm{Z}$.
They satisfy the anticommutation relations
$\{a_i,a^\dagger_j\}=\delta_{ij}\,,\{a_i,a_j\}=0=\{a^\dagger_i,a^\dagger_j\}$.
In (\ref{a}) we have taken already the limit $N\to\infty$. The transfer matrix is $e^{-t H_F}$, $t\geq 0$, and in the limit $T\to\infty$ one has to compute the ground state expectations for $H_F$. A macroscopic facet emerges as $\lambda \to \infty$.

In Fig. \ref{fig2} we display a typical ledge configuration for the TLK model with volume constraint. There is no further ledge above the one shown and for $j \to -\infty$ ledges are perfectly flat and densely packed.

Since a ledge corresponds to a fermionic world line, the average step
density $\langle \eta_j(t)\rangle_\lambda = \rho_\lambda(j)$ is
independent of $t$ and given by $\langle \eta_j(t)\rangle_\lambda
= \langle a_j^\dagger a_j\rangle_\lambda$ with $\langle
\cdot \rangle_\lambda$
on the right denoting the ground state expectation
for $H_F$. By the linear potential
in (\ref{a}) steps are suppressed for large $j$. Hence the average surface
height $h^\lambda_j(t)$ at $(j,t)$, relative to the high symmetry plane, equals
\begin{equation}{\label{height}}
h_j^\lambda(t)=-\sum_{k=j}^\infty \langle \eta_k(t)\rangle_\lambda\, .
\end{equation}
$\langle a_j^\dagger a_j\rangle_\lambda$ can be computed in terms of
the Bessel function $J_j(z)$ of integer order $j$ and its
derivative $L_j(z)=(\textrm{d}/\textrm{d}j)J_j(t)$ with the result
\begin{eqnarray}\label{b}
\rho_\lambda(j)=\langle
a^\dagger_ja_j\rangle_\lambda&=&\lambda\big(L_{j-1+
2[\lambda]}(2\lambda)J_{j+2[\lambda]}(2\lambda) \\ & &
-L_{j+2[\lambda]}(2\lambda)J_{j-1+2[\lambda]}(2\lambda)\big) \nonumber
\end{eqnarray}
where $[\cdot]$ denotes the integer part. For large $\lambda$
the height $h_j^\lambda(t)$ is of order $\lambda$. Therefore
we rescale the lattice spacing by $1/\lambda$.
Then $\lim_{\lambda \to \infty}\lambda^{-1}h_{[\lambda r]}(\lambda t)
= h_{\textrm{eq}}(r,t)$ with the macroscopic equilibrium crystal shape
\begin{eqnarray}\label{c}
h_{\mathrm{eq}}(r-2,t)=\left\{
\begin{array}{l}
 r \hspace{30pt}\mbox{for $r\leq -2$\,,}\\[6pt]
    \frac{1}{\pi}\big(r\arccos(r/2)-\sqrt{4-r^2}\big)\\
\phantom{0} \hspace{30pt} \mbox{for $-2\leq r \leq 2\,$,}\\[6pt]
0\hspace{30pt} \mbox{for $r\geq 2\,$.}
      \end{array}\right.
\end{eqnarray}
Thus under volume constraint the TLK model has two facets, one with
slope $1$, the other one with slope $0$, joined by a rounded piece. The upper facet
edge is located at $r=0$. It has zero curvature. Expanding near $r=0$ results in
$h_{\mathrm{eq}}(r,t) \cong -\frac2{3\pi} (-r)^{3/2}$, consistent with
the Pokrovsky-Talapov law.

With the exact result (\ref{b}) it becomes possible to refine the resolution.
The appropriate step size is $\lambda^{1/3}$ lattice constants. For the step
density $\rho_\lambda(j)=\langle a_j^\dagger a_j\rangle_\lambda$ close to $r=0$ one finds
\begin{equation}\label{d}
\lim_{\lambda\to\infty}\lambda^{1/3}\rho_\lambda(\lambda^{1/3}x)
=-x{\mathrm{Ai}}(x)^2+{\mathrm{Ai}}'(x)^2\,,
\end{equation}
$\mathrm{Ai}$ the Airy function. (\ref{d}) has the asymptotics
\begin{eqnarray}\label{e}
& & \frac{1}{\pi}\sqrt{|x|} \quad \mathrm{for} \quad x\to - \infty\,,
\\ & & \frac{1}{8 \pi x}\exp(-4x^{3/2}/3) \quad \mathrm{for}\quad x\to
\infty\,. \nonumber
\end{eqnarray}

Our real interest are the border ledge fluctuations. Clearly the
border ledge is the top fermionic world line which we denote
by $b_\lambda(t)$. $b_\lambda(t)$ takes integer values and is piecewise
constant with unit size kinks. Since, at fixed $t$, the steps in the bulk
have approximately the same statistics as a GUE random matrix, one would
expect that the transverse fluctuations of the border ledge equal those
of the largest eigenvalue. Indeed, using
the fermionic transfer matrix combined with an asymptotic analysis \cite{PS},
one finds that
\begin{equation}\label{g}
\lim_{\lambda\to\infty}
{\mathrm{Prob}}(\{b_\lambda(0)\leq\lambda^{1/3}a\})=F_2(a),\quad a\in\mathbbm{R}\, .
\end{equation}
In the random matrix community $F_2(a)$ is known as the Tracy-Widom distribution~\cite{TW}. The corresponding probability density $\textrm{d}F_2(a)/\textrm{d}a$ has an upper tail as $\exp(-\frac43 a^{3/2})$ and a lower tail as $\exp(-\frac{1}{12}|a|^3)$.

In our context an experimentally more accessible quantity is the ledge wandering
$\langle[b_\lambda(t)-b_\lambda(0)]^2\rangle$. In the limit of
large $\lambda$ it has been computed in \cite{PS} with the result
\begin{equation}\label{j}
\text{Var}\big(b_\lambda(t)-b_\lambda(0)\big)\cong\lambda^{2/3}g(\lambda^{-2/3}t),
\end{equation}
using the short-hand $\text{Var}(X)=\langle(X-\langle X\rangle)^2\rangle$.
Thus the transverse fluctuations are on the scale $\lambda^{2/3}$.
For small $s$ the scaling function $g(s)$ is linear in $s$,
$g(s)\simeq 2|s|$, indicating that for small, on the scale
$\lambda^{2/3}$, separations the border ledge has random walk statistics.
On the other hand $g(s)$ saturates for large $s$, $g(s)\simeq
g(\infty)-c/s^2$, reflecting that the border ledge fluctuations
are stationary (on the scale $\lambda^{2/3}$). For the leading term one finds
$g(\infty)=\lim_{\lambda\to\infty}\lambda^{-2/3}2 \langle b_\lambda(0)^2\rangle
= 1.6264$. The subleading coefficient $c$ has recently been derived in \cite{AM,Wid}
with the result $c=2$.

Within the volume-constrained TLK model we arrived at an interesting prediction for the
border ledge fluctuations. To be convincing we have to check against a more realistic
model, for which we take the three-dimensional Ising model at low temperatures. At fixed
crystal volume the equilibrium shape is then a cube with rounded corners. Taken literally
this model is still too complicated and we simplify through an SOS-type approximation by
allowing only atomic configurations which are lattice convex. This means
that, when the crystal is
cut along any line parallel to the major axes, then the atoms fill a single interval
(no holes). We use translation invariance to choose our coordinate system in such
a way that the crystal lies in the
positive octant of $\mathbb{Z}^3$ with three of the
facets coinciding with parts of the planes spanned by the three coordinate axes. If one
now restricts attention to the piece of the crystal close to the origin, then the actual
crystal shape can be represented by a height function $h(i,j)$, where $i$, resp. $j$,
refers to the (100), resp. (010), axis. By construction\\
{\it (i)}:$\,\,\, h(i,j) \geq 0$\,,\\
{\it (ii)}: $\,h(i,j) \geq h(i+1,j)\,, \quad h(i,j)\geq h(i,j+1)$\,.\\
The number of atoms missing, relative to the perfect cube, is
\begin{equation}
V(h) = \sum _{i,j \geq 0} h(i,j)\,.
\end{equation}
Thus the volume constraint translates into\\
{\it (iii)}: $\,\,V(h) = const\,.$\\
Every atomic configuration satisfying $(i) - (iii)$ has the same number of broken bonds and
thus the same energy. Therefore our simplified version of the 3D Ising equilibrium
droplet is to allow only atomic configurations which have a height
function satisfying $(i) - (iii)$ and to give them equal statistical weight.
Note that our model is purely
entropic. Fig. \ref{fig1} shows a typical sample with $V(h) = 3 \times 10^5$.

By projecting along the (111) direction the Ising corner model is equivalent to
tilings of the plane with rhombi of three distinct orientations. In this version
the surface tension is computed in \cite{Wu}.
According to the Andreev construction
the Legendre transform of the surface free energy yields the equilibrium crystal
shape \cite{A,BH}. Convenient formulas are available in \cite{CK}, where it is also
established that, for the constraint $V(h) = N$, in the limit
$N \to \infty$ with corresponding lattice spacing $N^{-1/3}$ the
equilibrium crystal shape is attained with probability one. From the implicit formula
for the shape it can be deduced that near facet edges the Pokrovsky-Talapov law
holds. The facet edge can be computed explicitly. If we consider the facet which lies in the (001) plane and denote the coordinate along (100) by $\tau$ and the one along (010) by $b_{\infty}$, then the macroscopic facet edge is given by
$b_\infty(\tau)=-\ln(1-e^{-\tau})$, $\tau > 0$.

As observed in \cite{OR}, the 3D Ising corner can be analyzed through fermionic
techniques. In particular, one can study the border ledge fluctuations. The details are rather intricate and given elsewhere
\cite{FS}. Here we only report on those results which allow us to gain some
understanding of the universal properties of ledge fluctuations.
We introduce the scaling parameter $\ell$ by
$N=\frac14\zeta(3)\ell^3$, $\zeta(3)=1.202\dots$ being Apery's
constant. The atomic border ledge position in the (001) plane is
given by $b_\ell(x)$, $x = 0,1,2,\ldots\,$.  $b_\ell(x)$ takes
positive integer values and is decreasing as $b_\ell(x+1) \leq
b_\ell(x)$. The ledge has only South and East turns and meanders
close to its asymptotic mean $\ell b_\infty(x/\ell)$.  We zoom at the
fixed macroscopic edge point $(\ell \tau,\ell b_{\infty}(\tau))$, $
\tau >0$. Upon proper rescaling \cite{FS}, one recovers exactly the
same statistics as in (\ref{g}). More precisely, for large $\ell$,
\begin{equation}\label{l}
  \text{Var}\big(b_\ell(\ell \tau +x)-b_\ell(\ell \tau)\big)\cong
  \left(\tfrac12 A\ell\right)^{2/3} g\Big(\frac{A^{1/3}}{2^{1/3}\ell^{2/3}}x\Big),
\end{equation}
with $A=b_\infty''(\tau)$.  (\ref{l}) differs from (\ref{j}) in two
respects. Firstly, to obtain the border ledge fluctuations one has to
subtract the systematic mean. Since for our particular model the
macroscopic facet edge is explicit, the subtraction is $\langle
b_\ell(\ell \tau +x)\rangle - \langle b_\ell(\ell \tau)\rangle \cong
b_{\infty}'(\tau) x + \frac12 b_\infty''(\tau) \ell^{-1} x^{2}$ with
negligible higher order corrections. Secondly, model-dependent properties
enter indirectly through the coefficient $A$. Since $g(s) = 2|s|$ for small $|s|$,
$\text{Var}\big(b_\ell(\ell \tau +x) -b_\ell(\ell \tau)\big)=A|x|$,
$A$ can be identified with the local wandering, resp.\ diffusion, coefficient.

The border ledge of the TLK model and the 3D Ising corner have the same
scaling behavior, which suggests the scaling to hold in greater generality.
To obtain the form which properly distinguishes between model-dependent and universal properties we have to rely on a few notions from the thermodynamics of equilibrium crystal shapes \cite{AA}.
Let us denote by $h(x,y)$ the height of a vicinal surface relative to the
high symmetry reference plane. We find it convenient to measure $h$ in number of atomic layers, whereas $x,y$ are measured in a suitable macroscopic unit. Thus $h$ is dimensionless and $x,y$ have the dimension $[length]$. Further let
$k_{B}T f({\bf u})$ be the surface free energy per unit projected area depending on the local slope ${\bf u}=\nabla h$. Below the roughening
transition $f$ has a cone at ${\bf u}=0$ and for small ${\bf u}$ behaves as
\begin{equation}\label{m}
f({\bf u})\cong \gamma(\theta) |{\bf u}| + B(\theta) |{\bf u}|^3
\end{equation}
with $\theta$ the polar angle of ${\bf u}$ \cite{GM}. The line
stiffness $\widetilde \gamma$ is defined through
$\widetilde\gamma(\theta) = \gamma(\theta)+\gamma''(\theta)$. As
argued in~\cite{AAY}, for short range surface models the Gaussian
curvature of the equilibrium crystal shape has a universal jump across the facet edge, which implies the relation $\widetilde\gamma(\theta) B(\theta) = \pi^2/6$.
Let us denote by $\widehat f$ the Legendre transform of
$f$. If $\int dx dy f(\nabla h(x,y))$ is minimized under the
constraint of fixed volume, then the resulting equilibrium surface is
given by $h(x,y)=\ell \widehat f(\ell^{-1} x,\ell^{-1} y)$, where
$\ell$ is the Lagrange multiplier adjusted so to give the correct
volume. $h$ is convex downwards and has a convex facet lying in the
$x$-$y$ plane. The facet boundary is determined by $\gamma(\theta)$
alone. Close to the facet edge, $h\cong -\frac{2}{3} \gamma_{PT} d^{3/2}$
with $d$ the normal distance to the facet edge, which defines the Pokrovsky-Talapov coefficient $\gamma_{PT}$. Under Legendre transformation the angle $\theta$
becomes the angle between the $x$-axis and the outer normal to the
facet and, correspondingly, $\gamma_{PT}$, the local curvature
$\kappa$, and the distance $r$ of a point on the edge to the origin
are parametrized through this angle $\theta$. The relationship between
$\widetilde\gamma$ and $B$ implies
\begin{equation}\label{n}
\gamma_{PT}^2 \kappa = 2 \ell^{-2} \pi^{-2}.
\end{equation}

We return to the border ledge fluctuations close to a given angle
$\theta_0$. For this purpose it is convenient to center the $x$-$y$
axis coordinate system at $r(\theta_0)$ with the $x$-axis tangential
and the $y$-axis along the inner normal to the facet. In this frame,
we denote by $y=b(x)$ the fluctuating border
step. Then $\langle b(x) \rangle = \frac{1}{2} \kappa(\theta_0) x^2$,
in approximation. For sufficiently small $|x|$, still large on the
scale of the lattice, $b(x)$ is like a random walk and
$\textrm{Var}\big(b(x)-b(0)\big) \cong \sigma^2 |x|$, which defines
the local wandering coefficient $\sigma^2$. Following~\cite{AA} it is
natural to equate $\sigma(\theta)^2$ with the inverse stiffness
$\widetilde \gamma(\theta)^{-1}$.
This implies
\begin{equation}\label{o}
\sigma^2=\kappa \ell, \qquad \kappa=\pi^2\gamma_{PT}^2 \sigma^4/2
\end{equation}
valid for any point on the facet edge.

The general scaling form is obtained now by using the TLK model as benchmark.
Locally the border ledge performs a random walk with nearest neighbor hopping rate
1, see (\ref{a}), thus $\sigma^2 = 2$. From (\ref{c}) the PT coefficient is
$\gamma_{PT} = 1/\pi\sqrt{\ell}$ in our units. Using these two as model-dependent parameters yields the scaling form
\begin{equation}\label{p}
  \text{Var}\big(b(x)-b(0)\big)\cong (\pi \gamma_{PT})^{-4/3}
  g\big((\pi\gamma_{PT})^{4/3}\sigma^{2} x/2\big).
\end{equation}

Of course, through (\ref{n}), (\ref{o}), any other pair of
model-dependent parameter can be used to reexpress (\ref{p}).

As a control check, the Ising corner must also satisfy (\ref{p}).
This is indeed the case with coefficients $\kappa= \ell^{-1} \sigma^2$, $\sigma^2= b''_\infty(\tau)(1+b'_\infty(\tau)^2)^{-3/2}$, and \mbox{$\gamma_{PT}=2^{1/2} b''_\infty(\tau)^{-1} \ell^{-1/2} \pi^{-1} (1+b'_\infty(\tau)^2)^{3/4}$}.

To summarize, the border ledge of a facet has fluctuations of size $\ell^{1/3}$, thus much reduced in comparison with a simple random walk. We claim that the scaling form (\ref{p}) is universal within the class of surface models with short range interactions. The scaling function $g$ can be expressed through determinants of infinite dimensional matrices. Short and long distance behavior is known explicitly. In (\ref{p}) there are two material parameters. Once they are determined experimentally, the functional form of the variance for the ledge fluctuations follows.


\begin{thebibliography}{00}
\bibitem{SCV} S.~Surnev, P.~Coenen, B.~Voigtl\"ander, H.~P.~Bonzel,
and P.~Wynblatt, Phys. Rev. B {\bf 56}, 12131 (1997).
\bibitem{NBE} M.~Nowicki, C.~Bombis, A.~Emundts, H.~P.~Bonzel,
and P.~Wynblatt, Europhys. Lett.  {\bf 59}, 239 (2002).
\bibitem{NBEB} M.~Nowicki, C.~Bombis, A.~Emundts, and H.~P.~Bonzel,
preprint (2002).
\bibitem{PT} V.~L.~Pokrovsky and A.~L.~Talapov, Phys. Rev. Lett. {\bf 42}, 65 (1978).
\bibitem{Sp} H.~Spohn, J. Stat. Phys. {\bf 47}, 669 (1987).
\bibitem{SVR} D.~C.~Schl\"{o}{\ss}er, L.~K.~Verheij, G.~Rosenfeld,
and G.~Comsa, Phys. Rev. Lett. {\bf 82}, 3843 (1999).
\bibitem{KE} S.~V.~Khare and T.~L.~Einstein, Phys. Rev. B {\bf 54}, 11752
(1996).
\bibitem{KM} S.~Karlin and G.~McGregor, Pacific J. Math. {\bf 9}, 1141
  (1959).
\bibitem{D} F.~Dyson, J. Math. Phys. {\bf 3}, 1191 (1962).
\bibitem{EPL} T.~L.~Einstein and O.~Pierre-Louis, Surf. Sci. {\bf 424},
  L299 (1999).
\bibitem{ERC} T.~L.~Einstein, H.~L.~Richards, S.~D.~Cohen, and O.~Pierre-Louis, Surf. Sci. {\bf 493}, 460 (2001).
\bibitem{BV} J.~Villain and P.~Bak, J. Physique {\bf 42}, 657 (1982).
\bibitem{PS} M.~Pr\"ahofer and H.~Spohn, J. Stat. Phys. {\bf 108}, 1071
(2002).
\bibitem{TW} C.~A.~Tracy and H.~Widom, Commun. Math. Phys. {\bf 159}, 151 (1994).
\bibitem{AM} M.~Adler and P.~van Moerbeke, preprint, arXiv:math.PR/0302329.
\bibitem{Wid} H.~Widom, preprint,  arXiv:math.PR/0308157 (2003).
\bibitem{Wu} F.~Y.~Wu, Phys. Rev. {\bf 168}, 539 (1968).
\bibitem{A} A.~F.~Andreev, Zh. Eksp. Teor. Fiz. {\bf 80} 2042 (1981); Sov. Phys.-JETP {\bf 53}, 1063 (1982).
\bibitem{BH} H.~W.~J.~Bl\"ote and H.~J.~Hilhorst, J. Phys. A: Math. Gen. {\bf 15}, L631 (1982).
\bibitem{CK} R.~Cerf and R.~Kenyon, Comm. Math. Phys. {\bf 222}, 147 (2002).
\bibitem{OR} A.~Okounkov and N.~Reshetikhin, J. Amer. Math. Soc. {\bf 16} 581 (2003).
\bibitem{FS} P.~L.~Ferrari and H.~Spohn, J. Stat. Phys. {\bf 113}, 1 (2003).
\bibitem{AA} Y.~Akutsu and N.~Akutsu, J. Phys. A: Math. Gen. {\bf 19}, 2813 (1986).
\bibitem{GM} E.~E.~Gruber, W.~W.~Mullins, J. Phys. Chem. Solids {\bf 28}, 875 (1967).
\bibitem{AAY} Y.~Akutsu, N.~Akutsu, and T.~Yamamoto, Phys. Rev. Lett. {\bf 61}, 424 (1988).
\end{thebibliography}
\end{document}